\documentstyle[12pt]{article}
\newcommand{\be}{\begin{eqnarray}}
\newcommand{\ee}{\end{eqnarray}}

\newcommand{\simgreater}{\buildrel > \over \sim}

\begin{document}
\begin{flushright}
{\small { WSU--NP--93--3 }}
\end{flushright}
\bigskip 
\thispagestyle{empty}
\begin{center}
{\large  {\bf TWO PION BOUND STATES IN A HOT PION GAS}}\\
  \bigskip
  \bigskip
{\large G. Chanfray,$^{a}$ P. Schuck,$^{b}$ and G. Welke$^c$}\\
\bigskip
{\small $^a$IPN, Universit\'e Claude Bernard Lyon I, 43 Bd du 11
Novembre 1918,\\
\vskip -5pt
F--69622 Villeurbanne C\'edex, France.\\
$^b$ISN, IN2P3--CNRS/Universit\'e Joseph Fourier, 53 avenue des
Martyrs,\\
\vskip -5pt
F--38026 Grenoble C\'edex, France.\\
$^c$Department of Physics and Astronomy, Wayne State University,\\
\vskip -5pt
Detroit, MI--48202, U.S.A.}\\
\bigskip
\end{center}
\bigskip
\bigskip
\begin{center}
\noindent {\large {\bf Abstract}}
\end{center}
The occurrence of poles in the $2\pi$ propagator in a 
hot pion medium is studied. The domain of non--trivial
solutions to the corresponding bosonic gap equation is investigated as
a function of the chemical potential and temperature of the gas.

\bigskip
\bigskip
\vfill
\noindent {\bf PACS Indices:}~ ~25.75.+r~ ~25.80.Dj~ ~11.10.St

\newpage     

When two heavy nuclei collide at very high energy, the matter that is
created differs qualitatively from that traditionally studied in nuclear and
elementary particle physics. In the initial stages of the collision, copious
production of gluons and quarks in a large volume leads to a rapid increase in
the entropy, and the distinct possibility of a new phase of matter
characterized by deconfined partonic degrees of freedom. One therefore hopes
that relativistic heavy ion experiments at Brookhaven and CERN will
provide some insight into the structure of the QCD vacuum,
deconfinement, and chiral symmetry restoration.

The hot transient system eventually evolves into a gas of
hadrons at high energy densities, whose interactions may well obscure
many potential signals of the deconfined phase.
To study this problem, extensive use has been made of sequential scattering
models, such as hadronic cascades \cite{RQMD,ARC,WB}. In principle, these
trace the evolution of the system from hadronization to freeze--out,
and can assign features in the data to the very early dense phase. 
For example, calculations using the Boltzmann equation with 
bose statistics to simulate heavy ion collisions at beam
energies of $200~A\cdot{\rm GeV}$ \cite{WB,GR} show that if the 
initial pion density is
large enough, the low transverse momentum component of their spectra will be
enhanced during the evolution of the pion gas \cite{NA35}.
Implicit in these studies is a thermalization rate governed by free
space $\pi$--$\pi$ interactions. However, recent investigations
\cite{ACSW,BBDS} of the effect
of finite phase space occupation on intermediate states suggest that
the t--matrix may well be suppressed within the environment of
a relativistic heavy ion collision (RHIC) \cite{BDSW}. This quantitatively
affects the conclusions one may draw from such simulations, and points directly
to the need to better understand hadronic properties and interactions
in RHICs.

Here we wish to reconsider $\pi$--$\pi$ interactions in the presence
of a dense and hot pion gas along the lines of our previous approach 
\cite{ACSW}. We shall address in some detail the question of pion pair
formation in the medium. As we shall see, the in--medium $2\pi$
propagator exhibits a pole between a lower and an upper critical
temperature, signaling a possible instability with respect to pion
pair formation.

To simplify the solution of the $2\pi$ propagator, we consider
a separable model for the free space $\pi$--$\pi$ interaction
\cite{JL}, in which the $2\pi$ state couples to an
intermediate $\sigma$ ($\ell\!=\!0,I\!=\!0$) or $\rho$
($\ell\!=\!1,I\!=\!1$) meson.
The corresponding scattering matrix may be derived from the following
effective Hamiltonian
\be
K &=& H \;-\; \mu \:(N_\pi \;-\; 2\, N_\alpha) \nonumber \\
&=& \sum_k \: (\omega_k-\mu) 
\:b_k^\dagger b_k \;+\; \sum_a\: (\Omega_a-2\mu)\:
\sigma_a^\dagger \sigma_a \nonumber \\
&~&~ ~ ~ ~ ~+\; \frac 12 \,\sum_{k_1 k_2 a} \,\bigg\{\,b_{k_1} b_{k_2}
\, (\sigma^\dagger_a+
\sigma_{-a}) \: \langle a\! \mid W \mid\! k_1 k_2 \rangle
\;+\; {\rm h.c.} \bigg \} ~ ~,\label{ham2}
\ee
\noindent where the $b_k$ ($b_k^\dagger$) are the pion annihilation (creation)
operators labeled
by momentum ${\vec k}$ with energy $\omega_k=\sqrt{{\vec
k}^{\,2}\!+\! m_\pi^2}$. The $\sigma_a$ ($\sigma_{-a}^\dagger$) 
are the corresponding  operators for the intermediate meson labeled
by $a=(\alpha,{\vec P})$ ($-a=(\alpha,-{\vec P})$), where
$\alpha=(\ell,I)$ are the spin--isospin quantum numbers. The state's
energy is $\Omega_a=\sqrt{{\vec P}^{\,2}\!+\!M_\alpha^2}$, where 
${\vec P}={\vec k}_1+{\vec k}_2$ is the total momentum, and $M_\alpha$
the mass of the exchanged meson. We chose a Yukawa coupling
\be
\langle a\! \mid W \mid\! k_1 k_2 \rangle &=& \sqrt{2}\;
(2\Omega_a 2\omega_{k_1} 2\omega_{k_2})^{-1/2}\; V_\alpha(k)\label{couple} \\
V_\alpha(k) &=& 4\pi \: g_\alpha \: \omega_{k}
\sqrt{M_\alpha} \: v_\alpha(k) \label{potform} \\
v_\alpha(k) &=& \frac{(k/k_\alpha)^{\ell}}{[
1+(k/k_\alpha)^2]^{\ell+1}}~ ~ ~,\label{formff}
\ee
\noindent where $k$ is the relative momentum of a pion in the c.m.
frame of the pion pair with total momentum ${\vec P}$. The constants
$g_\alpha$, $M_\alpha$ and $k_\alpha$ are chosen as in
Ref.~\cite{ACSW}, and thus our Hamiltonian reproduces the $\pi$--$\pi$
phase shifts in free space.

The introduction of a chemical potential in the effective Hamiltonian
may be justified as follows: There is considerable
evidence that the pion distribution at freeze-out in RHICs is near local
thermal equilibrium, but not in chemical equilibrium. Since the
$\pi$--$\pi$ interaction is essentially elastic up to $1~{\rm GeV}$,
in the later stages of the evolution the inelastic collision rate is
insufficient to achieve full equilibrium. As the system
continues to expand toward freeze--out, the dominant number conserving
processes still thermalize the system, but a non--zero chemical potential 
is built up \cite{WB,GR}.

At a formal level in a field theoretical scheme it is not possible to
define a total pion number operator from the canonical pion field. In
the context of our semi--relativistic effective Hamiltonian the total
pion number $\sum_i b^\dagger_{i{\vec k}} b_{i{\vec k}}$ ($i$ is the
isospin index) does not commute with $H$. It is, however, conserved on
time scales larger than the life--time of the intermediate meson: the
basic process is two pion annihilation to an intermediate $\sigma$ or $\rho$,
which subsequently decays back into two pions. In this spirit, we
introduce a chemical potential $\mu$ for the pion. The uncorrelated 
(causal) $2\pi$ propagator is then given by (at ${\vec P}=0$)
\be
G_{\pi\pi}^{\,T}(E,k) \;=\; \bigg \{
\frac{1}{E-2(\omega_k-\mu) + i\eta} \:-\: \frac{1}{E+2(\omega_k-\mu) + i\eta}
\bigg \}\;\coth \frac{\omega_k-\mu}{2T} ~ ~ ~.\label{medG}
\ee
\noindent Using the $K$--Hamiltonian Eq.~(\ref{ham2}),
we may rederive the in--medium $\sigma$--meson propagator,
which, after Fourier transformation, reads
\be
D_\alpha^{-1}(E) \:=\:  E^2 &-& (M_\alpha\!-\!2\mu)^2 \nonumber \\
&-& 2(M_\alpha\!-\!2\mu) 
\,\frac {1}{2}\int \!\frac{d^3k}{(2\pi)^3}  \, \mid\! \langle 0 \!\mid
W \mid \!
k,-k\rangle\! \mid^2 \,G^{\,T}_{\pi\pi}(E,k)~ .\label{medprop}
\ee
\noindent The pole condition $D^{-1}_\sigma(E)=0$ for the existence of
bound $2\pi$ states (with ${\vec P}=0$) 
in the sigma channel ($\ell=0,I=0$) follows from Eq.~(\ref{medprop})
with (\ref{couple}) and (\ref{medG}):
\be
\frac {1}{g^2_\sigma} &=& 2(M_\sigma-2\mu)\:
\int_0^\infty \! dk\:k^2 \:\frac{v_\sigma^2(k)}{E^2-(M_\sigma-2\mu)^2}
\; \frac {4(\omega_k-\mu)}{E^2-4(\omega_k-\mu)^2} 
\;\coth \frac {\omega_k-\mu}{2T} \nonumber\\
&\equiv& F_{\mu,T}(E)~ ~ ~.\label{tpole}
\ee
In Fig.~1 we show the function $F_{\mu,T}(E)$ for a fixed pion chemical
potential of $100~{\rm MeV}$, at five different temperatures (solid
lines). These curves intersect with $1/g_\sigma^2 \approx 486~{\rm MeV}$
(horizontal dashes) below energies of $2(m_\pi-\mu)$ (the bound state
domain) if the temperature lies between a lower and an upper critical
value of $T_0 \approx 80~{\rm MeV}$ ($E_{\rm pole}=2(m_\pi-\mu)$) and $T_c
\approx 137~{\rm MeV}$ ($E_{\rm pole}=0$), respectively.

Several observations can be made here: (i) One always obtains a pole
if the temperature lies between $T_0(\mu)$ (pole at
$E=2(m_\pi-\mu)$) and $T_c(\mu)$ (pole at $E=0$).
Thus, at fixed $\mu$, no matter how weak the 
potential $g_\sigma$ is (provided it is attractive in the neighborhood
of the $2\pi$ threshold), one always obtains a pole in a range of 
sufficiently high temperatures (for fermions at a sufficiently
low temperature). In practice, $T_0$ and $T_c$ will exceed sensible
values for pions as soon as $\mu$ drops below $\sim 100~{\rm MeV}$,
since they are increasing functions of $\mu$; (ii) 
for fixed $g_\sigma$, the pole position
shifts downward with increasing temperature (for fermions: pole position
moves up with increasing temperature); and (iii) the pole position
also shifts down with increasing $g_\sigma$ (as for fermions).
As a function of temperature, we therefore see a behavior for bosons
opposite to that for fermions. 

The fact that increasing temperature
reinforces the binding is somewhat counterintuitive, but it is an
immediate consequence of the coth factor associated with bose
statistics in Eq.~(\ref{tpole}) (fermions: tanh), which enhances the
medium density, effectively reinforcing the 2--body matrix element 
for increasing temperature. However, the bound state energy drops to zero
at a certain critical temperature $T_c(\mu)$. For $T>T_c$, a real
solution ceases to exist. As usual, such a zero eigenvalue signals a
phase transition point. We shall see shortly that for $T\simgreater T_c$ the
system has become so dense that pairs (corresponding to the pole)
begin to obstruct each other and dissolve in the medium.

In analogy with the Cooper pole in fermion scattering, a pole in the
two boson propagator also implies a non--trivial solution to the 
corresponding
``gap equation.'' It is straightforward to derive
such an equation from the Hamiltonian Eq.~(\ref{ham2})
\be
\Delta_k \;=\;  8\pi^2 g_\sigma^2 \: \int \! 
\frac{d^3q}{(2\pi)^3} \; \frac{v_\sigma(k)\, v_\sigma(q) }{M_\sigma-2\mu} 
\; \frac {\Delta_q}{2E_q} \; \coth
\frac{E_q}{2T}~ ~ ~.\label{gapeqn}
\ee
\noindent where $v_\sigma$ is given by Eq.~(\ref{formff}), and 
the boson quasiparticle energy is
$E_k^2 = (\omega_k-\mu)^2 - \Delta_k^2$. The ansatz $\Delta_k=\delta\, 
v_\sigma(k)$ reduces Eq.~(\ref{gapeqn}) to a non--linear equation for the 
gap strength $\delta$.

In spite of the formal similarities of Eq.~(\ref{gapeqn}) with the
corresponding fermionic gap equation, there are important differences: 
For bosons, the $\Delta_k^2$ is subtracted in $E_k$ (fermions: 
added), and the temperature factor is a hyperbolic cotangent
(fermions: tanh). 
We have seen previously that at sufficiently high
temperature ($>T_0(\mu)$) there exists a pole in the $2\pi$
propagator, {\it i.e.},
finite $T$ favors binding, contrary to the fermion case. As we shall
see, a similar feature occurs for the gap. At fixed $\mu$, above  a 
temperature $T^\prime_0(\mu)$ a finite gap appears in the spectrum, 
which disappears
again beyond a critical temperature which will turn out to be
$T_c(\mu)$. At temperatures larger than $T_c$, the medium density increases
to the point where the pairs dissolve ({\it i.e.} $\delta=0$).

Let us now study the domain of existence of a real, finite value of $\Delta$
in the $\mu$--$T$ plane. Rewriting the gap equation (\ref{gapeqn}) as
\be
\frac {1}{g^2_\sigma} &=& 2\: \int_0^\infty \! dk\,k^2
\:\frac{v_\sigma^2(k)}
{M_\sigma-2\mu} \; \frac {1}{\sqrt{(\omega_k\!-\!\mu)^2-\delta^2
v_\sigma^2(k)}}\;\coth \frac {\sqrt{(\omega_k\!-\!\mu)^2-\delta^2
v_\sigma^2(k)}}{2T} \nonumber\\
&\equiv& G_{\mu,T}(\delta)~ ~ ~,\label{inteqn}
\ee
\noindent it is clear that for fixed $\mu$ and $T$, $G_{\mu,T}(\delta)$
increases monotonically from $\delta\!=\!0$ to 
$\delta_{\rm max}\!=\!m_\pi\!-\!\mu$. Further,
for fixed $\mu$, $G_{\mu,T}$ also increases with temperature. This behavior
is shown in Fig.~2 for $\mu\!=\!100~{\rm MeV}$ and various values of
$T$ (solid lines). The pair potential $\delta$ is given by the 
intersection of the solid lines with the horizontal dashed line. With
decreasing temperature, it varies from zero at $T\!=\!T_c$ to 
$\delta_{\rm max}\!\equiv\! m_\pi\!-\!\mu$ at $T\!=\!T^\prime_0$. 
In Fig.~3 we show the gap strength $\delta=\Delta_k/v_\sigma(k)$ as a
function of $m_\pi-\mu$. For a given temperature, $\delta$ vanishes 
at some upper critical chemical potential, and increases to $\delta_{\rm
max}$ at a lower critical chemical potential. 

The critical temperature $T_c$ for which
the gap vanishes is determined by linearizing Eq.~(\ref{gapeqn}), {\it
i.e.}, setting $\delta\!=\!0$ in Eq.~(\ref{inteqn}).
The resulting  equation is identical to the 
pole equation (\ref{tpole}) at $E=0$, and thus the gap and
the bound state both disappear at the same critical temperature $T_c$.
Eq.~(\ref{tpole}) is equivalent to what is known in the fermion case
as the particle--particle RPA. Thus we see that in the
boson case, too, one has the well known property that a phase transition to
a pair condensate is signaled by the appearance of a zero eigenvalue
in the RPA eigenfrequencies (Thouless criterion) \cite{RS8}.

For fixed $\mu$, as $T$ decreases, $\delta$ increases (see Fig.~2),
until it reaches a maximum value of $m_\pi\!-\!\mu$ at
$T\!=\!T_0^\prime$. At this point $E_{k=0}\!=\!0$, and the gap in the
excitation spectrum has ceased to exist. Any further
lowering of the temperature would yield a purely imaginary
quasiparticle energy. In order to understand what happens, we
investigate the pion momentum distribution
\be
n_k \;=\; \langle b_k^\dagger b_k \rangle 
\;=\;  \frac{1}{\exp (E_k/T) \:-\: 1} 
\;+\; v_k^2 \: \coth \frac{E_k}{2T} \label{pidist}
\ee
\noindent where $v_k^2 = [\,(\omega_k-\mu)/(2E_k) - 1/2\, ]$.
Since $E_{k=0} \rightarrow 0$ for $T\rightarrow T_0^\prime$, we see
that $n_{k=0}$ diverges at $T=T_0^\prime$. The quasiparticles
start to condense in the ${\vec k}=0$ state in very much the same way
as in the unpaired state when real particles condense as the chemical
potential approaches $m_\pi$ for decreasing temperature. The interesting
point in the paired case is that the condensation of quasiparticles
happens at $\mu\!<\!m_\pi$, since it is counterbalanced by the finite
value of $\delta$.

The region between the solid lines in Fig.~4 summarizes
the domain of non--vanishing $\delta$ in the $\mu$--$T$ plane.
We see that the temperature $T_0^\prime$ -- where the gap closes up 
the excitation spectrum to zero at $k\!=\!0$, and quasiparticles 
begin to condense as singles --
is very close to the $T_0$ (dot--dashed line)
where the pole in the $2\pi$ propagator
ceases to exist. The entire $\mu$--$T$ domain of non--trivial 
solutions to the gap equation is therefore intimately linked to the
region where a pole exists in the $2\pi$ propagator.

In summary, we have shown that finite temperature induces real poles
in the $2\pi$ propagator, even for situations where there is no $2\pi$
bound state in free space. The situation is analogous to the
Cooper pole of fermion systems, and we therefore studied the
corresponding bosonic ``gap'' equation. This equation has non--trivial
solutions in a certain domain of the $\mu$--$T$ plane. 
Such a region always exists, even in the limit of infinitesimally weak
attraction. This is different from the $T=0$ case discussed by Saint
James and Nozi{\`e}res \cite{StJNoz}, where a nontrivial solution to
the gap equation only exists when there is a two boson bound state in
free space. Our study has to be considered preliminary. The final aim
will be to obtain an equation of state for a hot pion gas within a
Bruckner--Hartree--Fock--Bogoliubov approach. One of the major
questions in this regard is how to avoid a collapse, since $\pi$--$\pi$
interactions are primarily attractive with no need for a hard core
repulsion down to very small distances. Also, the subtle question of
single boson versus pair condensation must be addressed (see Ref.~\cite{StJNoz}
and references therein). In view of the strongly
off--shell nature of the problem, the $\pi$--$\pi$
interaction should be chosen to be more realistic than is the case
here. Work on these topics is in progress.

We are grateful to G.~Bertsch, P.~Danielewicz and M.~Prakash for discussions. 
This work was supported in part by the U.S. Department of Energy 
under Grant No. DE-FG02-93ER40713.

\bigskip

\noindent {\Large {\bf Figure Captions}}

\noindent {\bf Fig.1}~ ~The function $F_{\mu,T}$ of Eq.~(\ref{inteqn})
versus energy, for a fixed chemical potential
of $\mu=100~{\rm MeV}$ and several temperatures (solid
lines). Intersections of the solid lines with the horizontal dashed line 
($1/g_\sigma^2$) correspond to poles in the
in--medium $\sigma$ channel $2\pi$ propagator. 

\noindent {\bf Fig.2}~ ~The function $G_{\mu,T}$ of Eq.~(\ref{inteqn}) versus
the gap strength $\delta$, for 
$\mu=100~{\rm MeV}$ and several temperatures (solid
lines). Intersections of the solid lines with the horizontal dashed line
($1/g_\sigma^2$) correspond to solutions of
the gap equation (\ref{gapeqn}). At the lower critical temperature
$T^\prime_0$ the quasiparticle energy vanishes ($\delta=\delta_{\rm max}$),
while the gap vanishes at the upper limit $T_c$.

\noindent {\bf Fig.3}~ ~The gap strength versus $m_\pi-\mu$ at
several temperatures.

\noindent {\bf Fig.4}~ ~The area between the solid lines 
corresponds to the domain in the
$\mu$--$T$ plane for which the gap equation (\ref{gapeqn}) has 
non--trivial solutions. The $2\pi$ propagator has a pole in the region
between $T_0$ (dot--dashed line, $E_{\rm pole}=2(m_\pi-\mu)$) and $T_c$
($E_{\rm pole}=0$).

\end{document}